\documentclass[%
 aip,
 amsmath,amssymb,
 reprint,%
]{revtex4-1}

\usepackage{graphicx}% Include figure files
\usepackage{dcolumn}% Align table columns on decimal point
\usepackage{bm}% bold math
\usepackage[utf8]{inputenc}
\usepackage[T1]{fontenc}

\usepackage{mathptmx} %cursive letters
\usepackage{lipsum}
\usepackage{color}

\usepackage{hyperref}

\begin{document}

\preprint{AIP/123-QED}

\title[]{Drop freezing: fine detection of contaminants  by measuring the tip angle}
% Force line breaks with \\

% Laboratoire de Physique des Solides, CNRS, Univ. Paris-Sud, Universit\'e Paris-Saclay, Orsay 91405, France

\author{François Boulogne}
 \email{francois.boulogne@u-psud.fr}
\author{Anniina Salonen}%
\affiliation{ 
Université Paris-Saclay, CNRS, Laboratoire de Physique des Solides, 91405, Orsay, France.%\\This line break forced with \textbackslash\textbackslash
}%

\date{\today}% It is always \today, today,

\begin{abstract}
           In this Letter, we show that the shape of a freezing drop of water can be sensitive to the presence of impurities.
            We measure the tip angle of water drops frozen on a cold plate.
            The fine changes in tip angle are robustly captured by our image analysis method, which shows a deviation from that of pure water in solutions with salt (NaCl), polymer (PEG) and surfactant (TTAB) starting at concentrations of $10^{-6}$, $10^{-4}$ and $10^{-6}$ M respectively.
            The method could be adapted into a portable water purity tester, but the work also highlights the complexity of water freezing as it is influenced by trace concentrations of impurities.
\end{abstract}

\maketitle

%%%%%%%%%%%%%%%%%%%%%%%%%%%%%%
%
% INTRODUCTION
%
%%%%%%%%%%%%%%%%%%%%%%%%%%%%%%

%\paragraph{Introduction}
Whereas a liquid drop deposited on a surface adopts the shape of a spherical cap, observations reveal that a frozen water droplet has a pointy shape \cite{Anderson1996,Schultz2001}, a topic that has been recently renewed by Snoeijer \textit{et al.} \cite{Snoeijer2012}.
The final morphology of frozen drops takes its roots in the lower density of ice compared to liquid water \cite{Schultz2001,Ajaev2003,Ajaev2004,Marin2014,Schetnikov2015} and is also affected by the contact angles \cite{Ajaev2003,Ajaev2004} at the tri-junction point, the point where ice, liquid, and vapor meet.
Mar\`in \textit{et al.} measured systematically the tip angle for various freezing temperatures and drop aspect ratios, and they concluded that the angle of the tip is universal \cite{Marin2014}.
The robustness of the tip angle has been also reported for asymmetric drops, \textit{i.e.} when the contact angle between the drop and the substrate is not uniform around the contact line \cite{Ismail2016}.

Applications of freezing processes are often oriented toward air conditioning systems, heat exchangers, wind turbines  \cite{Battisti2015} as well as aircraft \cite{Baumert2018} for which ice generates high maintenance costs worldwide.
Freezing is also heavily used in the food industry in freeze-drying processes to extract water and preserve food products \cite{Oetjen2004,Deville2006}.
The physics of ice is also naturally crucial in glaciology, in meteorology and in the climate change.

In this Letter, we investigate the effect of the presence of low concentrations of solutes on the final frozen drop morphologies.
As a comprehensive study on the variety of water solute would have been unrealistic, we choose an example of salt, polymer and surfactant, respectively.
Our measurements reveal that traces of these solutes lead to a measurable variation of the tip angle.
In particular, we show that the tip angle is more sensitive than conductivity or surface tension measurements.
Thus, we propose a method for assessing water purity by measuring the tip angle of a frozen drop from image analysis.

%%%%%%%%%%%%%%%%%%%%%%%%%%%%%
% Setup
%%%%%%%%%%%%%%%%%%%%%%%%%%%%%

\begin{figure}[h]
    \centering
    \includegraphics[width=\linewidth]{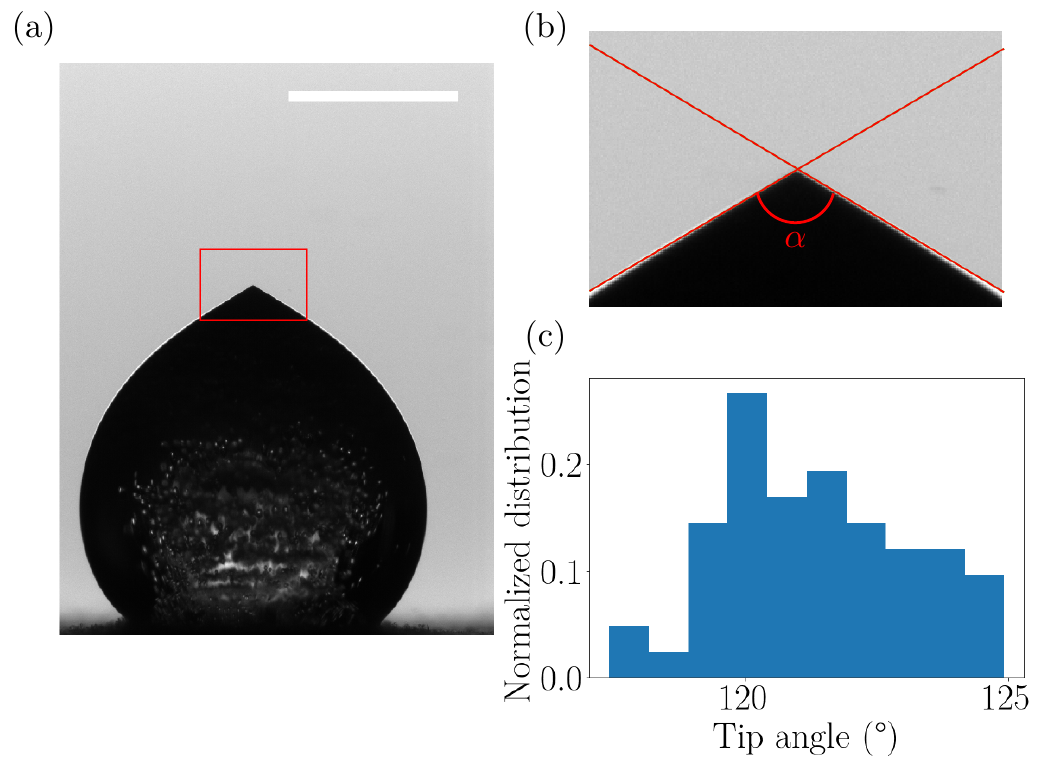}
    \caption{
    (a) Image of a 5 $\mu\ell$ pure water drop taken after the complete freezing on a Peltier plate cooled down at -18$^\circ$C.
    The red rectangle shows the region of interest centered on the tip, which is used for the tip angle measurement.
    The scale bar represents 1 mm.
    (b) Region of interest to extract the position of the interface. Red lines shows the straight lines detected by the Hough transform, which gives the tip angle $\alpha$.
    (c) Normalized distribution of the tip angles for pure water.
    }
    \label{fig:setup}
\end{figure}

%\paragraph{Methods}
Our experiments consist in recording the freezing of aqueous drops after deposition on a cold surface.
We use a Peltier module ($55\times 55$ mm$^2$, 80 W, Radiospare) %ref: 490-1418
to get a cold surface of controlled temperature set between -16~$^\circ$C and -19~$^\circ$C.
A water cooling system is used to extract the heat produced on the opposite side of the module.
The temperature is measured with a surface thermocouple of K-type (Radiospare). %ref: 342-8956
Direct observations are done with a horizontal microscope composed of a long working distance objective (2$\times$, Mitutoyo), a tube lens (1$\times$, Mitutoyo), and a video camera (CCD monochrome Manta, 1624$\times$1234 pixels, Allied Vision).

Once the surface reaches a temperature between -16~$^\circ$C and -19~$^\circ$C, a drop of 5~$\mu\ell$ is deposited directly on the cold surface of the Peltier module with a micropipette (Eppendorf, Research Plus).
Pure water on the surface of the Peltier module has a contact angle of about $90^\circ$ at room temperature, and the cold surface limits the drop spreading \cite{Herbaut2019}, which can result in a larger contact angle as shown on the pictures (Fig. \ref{fig:setup}a).
Experiments are performed at room temperature between 18~$^\circ$C and 20~$^\circ$C, which is also the initial temperature of the drop.
The drop starts to freeze at the contact of the cold surface and a freezing front propagates from the bottom of the drop to the top.
The microscope focus is adjusted during the drop freezing while images are recorded at a framerate of about 10 images per second.
For the analysis of the final shape, the picture is taken immediately after the tip formation, so that condensation and crystal growth on the tip have no influence on the morphological detection \cite{Snoeijer2012}.
In this Letter, we prepared solutions of sodium chloride (NaCl, purity $> 98\%$, Sigma-Aldrich) at $10^{-1}$ M,  polyethylene glycol (PEG, $M_w=300$,  Sigma-Aldrich) at $10^{-1}$~M, and tetradecyltrimethylammonium bromide (TTAB, purity $>99\%$, Sigma-Aldrich) at $4\times 10^{-3}$ M dissolved in pure water (Resistivity $18.2$ M$\Omega\cdot$cm, Purelab Chorus, Veolia).
Lower concentrations are obtained by diluting these solutions.
As we repeat the same deposition method, and that the drop volume and the surface temperature are kept constant in this study, the diameter of the drop in contact with the cold surface is $1.8\pm0.2$ mm, which is a weak variation.
For all the solutions presented in this Letter, the freezing duration of the drops are between 11 to 12 s, without effects of the nature of the solute or the concentration.

%%%%%%%%%%%%%%%%%%%%%%%%%%%%%
% Pure water
%%%%%%%%%%%%%%%%%%%%%%%%%%%%%

%\paragraph{Pure water}

Images are processed with the scikit-image library \cite{Vanderwalt2014} in Python.
The raw image, for which an example is shown in Fig.~\ref{fig:setup}(a), is thresholded to detect the position of the tip.
The tip, considered as the top pixel of the contour, is used to center a region of interest of 200 by 300 pixels (Fig.~\ref{fig:setup}(b)).
The frozen drop is surrounded by a white allow due to optical effects.
The drop edges are detected by considering the position of the outer part of the halo.
A line Hough transform is then performed to extract the two most prominent lines corresponding to each edge, which define the tip angle $\alpha$ represented in Fig.~\ref{fig:setup}(b).
We set a precision of $0.5^\circ$ for the detection of each edge in the Hough transform algorithm.
The implementation of this algorithm is provided in Supplementary Material.

First, we performed repeated measurements of the tip angle for pure water drops.
We measured an average angle $\alpha=121.1^\circ$ with a standard deviation of $\pm 1.8^\circ$ over a total of 54 drops.
The angle distribution is represented in Fig.~\ref{fig:setup}(c).
During the acquisition of the data for the solutions presented thereafter, we regularly performed experimental checks of the tip angle with pure water to ensure the reproducibility of this result.
Our measurements for solutions are averaged over 7 to 10 experiments.

In the study of Mar\`in \textit{et al.} in which a large number of tip angle measurements were conscientiously performed, the deposition method, the substrate, and the numerical approach to determine the angles is different from our method.
Previously reported angles for water drop are between $128^\circ$ and $155^\circ$ \cite{Snoeijer2012,Marin2014} and the method  has a precision of $\pm5^\circ$, whereas our method has a typical uncertainty of $\pm 2^\circ$.
The information necessary to compare the experimental procedure and the data analysis from former studies \cite{Marin2014,Ismail2016,Jambon-Puillet2019,Chu2019} is not sufficiently developed to analyze the minute  differences.
Therefore, we estimate that our observations are not contradictory with former studies and we will show that the precision reached by our algorithm makes amenable the measurement of fine physico-chemical effects.

%%%%%%%%%%%%%%%%%%%%%%%%%%%%%
% NaCl
%%%%%%%%%%%%%%%%%%%%%%%%%%%%%

\begin{figure}[htp]
    \centering
    \includegraphics[width=\linewidth]{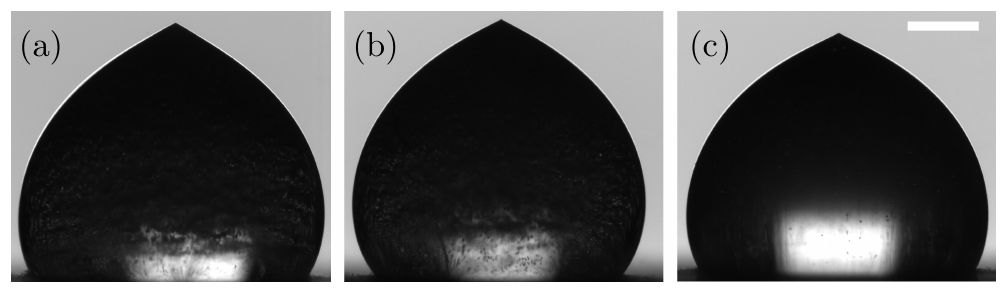}\\
    \includegraphics{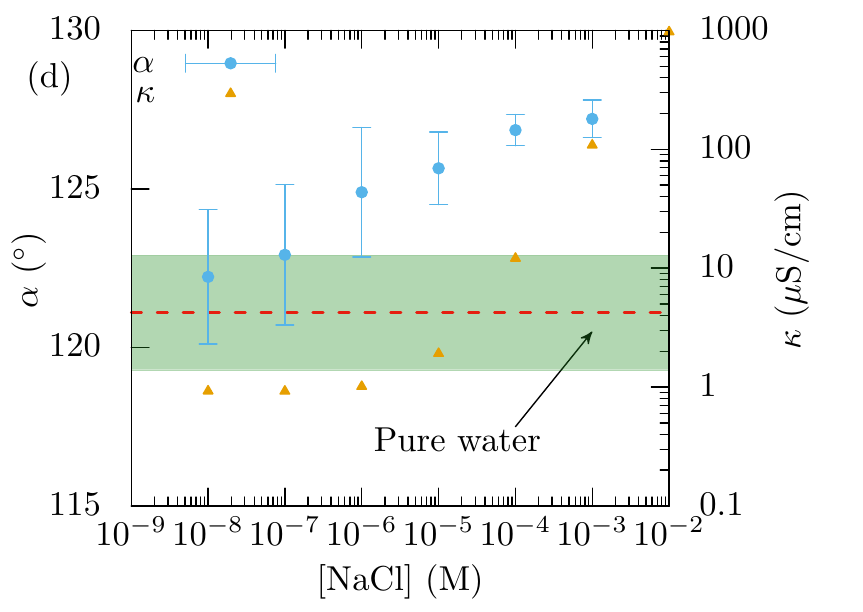}
    \caption{
    Images of the frozen drops of NaCl solutions at (a) $10^{-8}$ M, (b) $10^{-5}$ M, and (c) $10^{-3}$ M.
    The scale bar represents 0.5 mm.
    (d) Plot of the tip angle $\alpha$ and solution conductivity $\kappa$ at $20^\circ$C for NaCl solutions of different concentrations.
    The dashed red line is the angle value for pure water. The errorbars and the green colored band represent the standard deviation of the measurements.
    }
    \label{fig:nacl}
\end{figure}

%\paragraph{Salt}
After pure water, the second system we investigated is solutions of sodium chloride.
The pioneering observations on the freezing of salt solutions have been performed in the 17th century by melting ice from the sea water, which produces fresh water \cite{Nebbia1968}.
Indeed, ions have a low solubility in ice \cite{Petrenko1999,Vrbka2005} and ions concentrate in the liquid repealed by the freezing front.
Recent observations of the freezing of drops of salt solutions indicate the suppression of the tip effect and the rejection of a brine \cite{Singha2018}.
These experiments were performed at high salt concentrations of about 1 M, and the tip angles were not quantified.
Here, we focus on the freezing of dilute systems at concentrations below $10^{-3}$ M.

In Fig.~\ref{fig:nacl}, we show the tip angle $\alpha$ measurement for the range of NaCl concentrations between $10^{-8}$ and $10^{-3}$ M.
The tip angle continuously increases in this range with a typical variation of $5^\circ$.
At concentrations of about $10^{-2}$ M and above, we observed that brine is rejected at the tip of the drop as reported by Singha \textit{et al.} \cite{Singha2018}.
Down to concentrations of $10^{-6}$ M, effects of ions on the tip angle are clearly measurable.
We also measured the solutions' conductivity (InLab 741-ISM, Mettler Toledo), which are reported in Fig.~\ref{fig:nacl}.
Our measurements indicate a nearly constant value of the conductivity for salt concentrations below $10^{-6}$ M.
This effect is explained by the dissolution of the atmospheric CO$_2$ transformed in carbonic acid \cite{Light1995}, which screens the presence of sodium chloride in solution.
Interestingly, the measurement of the tip angle appears to be more sensitive than the conductivity regarding the presence of sodium and chloride ions, which indicates that the tip angle is less sensitive to carbonic acid.
Also, we noticed that at a concentration of $10^{-3}$ M, the opacity of the drop increases and the bubbles captured in the ice are less present.

%%%%%%%%%%%%%%%%%%%%%%%%%%%%%
% PEG300
%%%%%%%%%%%%%%%%%%%%%%%%%%%%%

\begin{figure}
    \centering
    \includegraphics{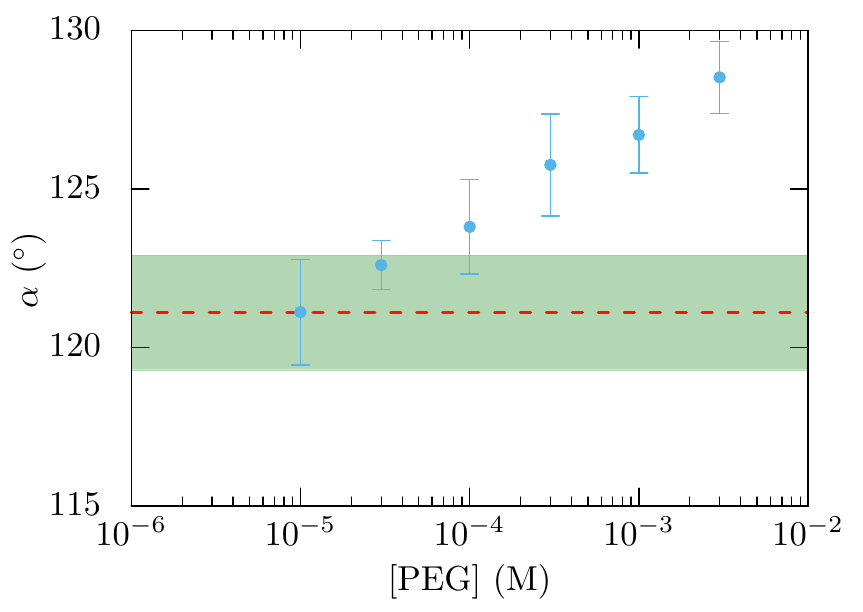}
    \caption{
     Tip angle $\alpha$ measured for different concentrations of PEG dissolved in pure water.
     The dashed red line is the angle value for pure water. The errorbars and the green colored band represent the standard deviation of the measurements.
    }
    \label{fig:peg300}
\end{figure}

%\paragraph{Polymer}
Next, we investigated a non-ionic system made of polymer (PEG) in solutions.
The measurements of the tip angle represented in Fig.~\ref{fig:peg300} indicate also an effect of the polymer concentration.
With this molecule, deviations from the tip angle for pure water can be assessed for concentrations typically larger than $10^{-4}$ M.
At a concentration of $10^{-2}$ M, we observed that the position of the frozen interface started to deviate significantly from a straight line, the tip has no longer a conical shape.
Therefore, we disregarded these large concentrations.
The drop shapes are very similar to those with NaCl, and again similarly, an increase of ice opacity has been observed.

%%%%%%%%%%%%%%%%%%%%%%%%%%%%%
% TTAB
%%%%%%%%%%%%%%%%%%%%%%%%%%%%%

\begin{figure}[ht]
    \centering
    \includegraphics[width=\linewidth]{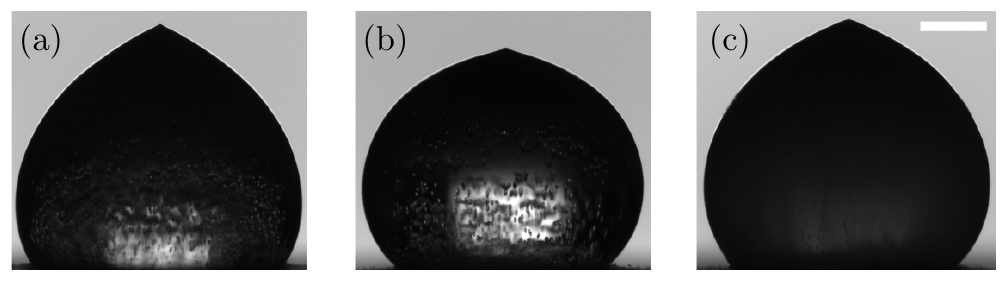}  \\
    \includegraphics{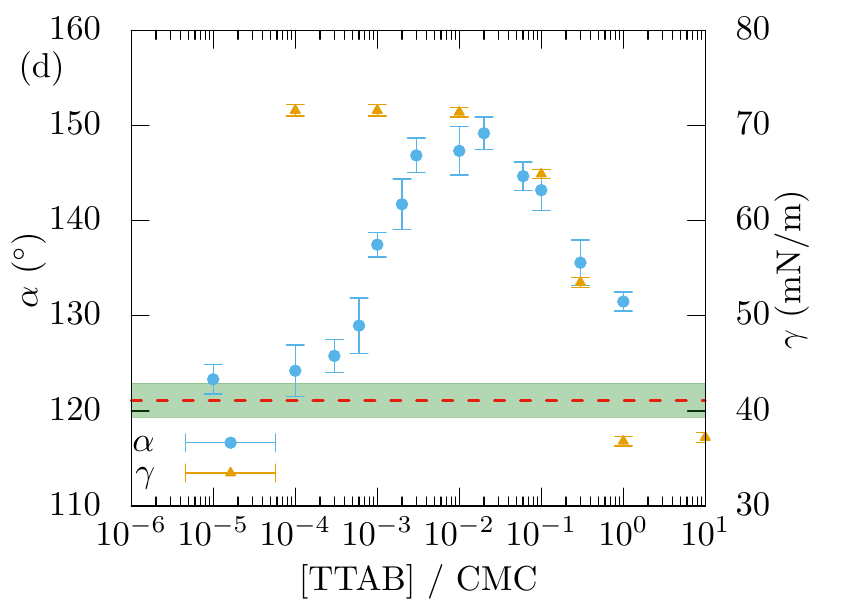}
    \caption{
    Images of frozen drops of surfactant solutions (TTAB) at (a) $1.0 \times 10^{-4}$ CMC,
    (b) $1.0 \times 10^{-2}$ CMC,
    and (c) $1.0$ CMC.
    The scale bar represents 0.5 mm.
    (d) Tip angle $\alpha$ as a function of the TTAB concentration normalized by the CMC (CMC = 4 mM\cite{Bergeron1997}).
    The dashed red line is the angle value for pure water. The errorbars and the green colored band represent the standard deviation of the measurements.
    We plot also the liquid-vapor surface  tension $\gamma$ at $20^\circ$C.
    At $10^{-3}$ CMC  and below, the measured surface tension is found to be identical to the one of water, at the precision of the measurements.
    }
    \label{fig:ttab}
\end{figure}

%\paragraph{Surfactant}
The last system studied in this Letter is solutions of TTAB.
Surfactant concentrations are commonly characterized by the critical micellar concentration denoted CMC.
We measured the surface tension by drop shape analysis using a rising bubble tensiometer (Tracker, Teclis, France).
The surface tension values shown in Fig.~\ref{fig:ttab} were taken after an equilibration time of 5000~s.
For TTAB, the CMC is 4 mM \cite{Bergeron1997}, a value confirmed by our surface tension measurements (Fig.~\ref{fig:ttab}).

In our freezing experiments and in contrast to the two former systems, we observed a non-monotonic variation of the shape of the drop and the resulting tip angle.
The tip angle reaches a maximum at a concentration of about $10^{-2}$ CMC and decreases afterwards.
As concentration increases, the ice becomes more turbid as illustrated in Fig.~\ref{fig:ttab}(a-c), as for the solutions of NaCl and PEG.
At a concentration of 10 CMC, we observe that the slope of the iced drop at the tip is not uniform anymore.
Therefore, we did not quantify the angle beyond this limit.

The surface tension curve presented in Fig.~\ref{fig:ttab}(d) shows that the presence of surfactant molecules is difficult to assess below $10^{-2}$ CMC.
Interestingly, the deviation of the tip angle $\alpha$ in the range $10^{-4}-10^{-2}$ CMC is clearly measurable.
We can also compare our method's sensitivity to the detection of traces of surfactants with Marangoni effects \cite{Arangalage2018}.
In their study, Arangalage \textit{et al.} show that an ethanol drop suspended above the interface of a probed solution leads to Marangoni flows able to detect the presence of TTAB molecules at concentrations larger than $10^{-3}$ CMC.
Our method is more sensitive in this case as surfactant concentrations are detected down to $10^{-4}$ CMC.

%%%%%%%%%%%%%%%%%%%%%%%%%%%%%
% Interpretation
%%%%%%%%%%%%%%%%%%%%%%%%%%%%%

%\paragraph{Discussion}

Particles and solutes can be either engulfed or rejected by a propagating freezing front. The behavior of solutes is particularly complex as it depends on the freezing rate, the physical and chemical properties of the particles, and their concentrations \cite{Halde1980,Dedovets2018,Jambon-Puillet2019}. Studies describing the interaction of solutes and a freezing front are scarce; the observations and the predictions under different conditions remain challenging. Although it is known that a slow freezing of salty water leads to ice of pure water \cite{Halde1980,Petrenko1999}, our protocol imposes a freezing front propagating typically at $100$ $\mu$m/s. The ice is made of multiple crystal domains entrapping impurities \cite{Barnes2004}. Our observations on the ice turbidity and the disappearance of bubbles corroborate a change of the ice composition.

Also, observations on concentrated solutions of salt indicate that a brine is rejected by the freezing front, as an ejection of liquid from the tip is observed \cite{Singha2018}.
We made similar observations for NaCl and PEG solutions, which indicates that the solute concentration increases in the liquid phase as the freezing front propagates. Therefore, the liquid composition is also changed during the freezing.

Former studies suggest that the frozen drop morphology depends on the material density and the angles at the tri-junction \cite{Ajaev2003,Ajaev2004}, and both are influenced by a changing drop composition. 
For the solutions studied in this Letter, the concentration of solutes is weak compared to the solvent. 
The tip angle is measured on approximately 1\% of the drop volume, and even if we assume a complete rejection of solutes before the freezing front, the material density change would not be significant.

Variations in the ice and liquid compositions must lead to variations of the interfacial energies and therefore to the angles at the tri-junction. 
The geometry of the tri-junction is out-of-equilibrium and depends on the surface tensions at liquid-ice, ice-vapor, and liquid-vapor interfaces.
The most striking effects on drop shape and tip angle occur with the surfactant solution, which also exhibits the most pronounced changes in interfacial energy with concentration, thus corroborating the importance of interfacial energies. However, confirmation of the exact changes requires explicit measurement of the tri-junction angles. This is complicated, both due to the geometry of the drop and its fast evolution. Further investigations with more advanced techniques are necessary to elucidate the precise effects of the solute chemistry on these properties. This study reveals that solutes may have significant effects of the frozen drop morphologies and that this discovery can be used to assess the purity of water.

%%%%%%%%%%%%%%%%%%%%%%%%%%%%%
% Conclusion
%%%%%%%%%%%%%%%%%%%%%%%%%%%%%

%\paragraph{Conclusion}
Our measurements show that the morphology of freezing water drops is significantly modified by the presence of impurities.
We propose an image analysis sensitive enough to capture the minute variation of the tip angle and therefore make possible the detection of low concentrations of solutes more sensitive than conductivity for sodium chloride, and more sensitive than surface tension measurements for TTAB solutions.
A test on tap water (nitrate: 19 mg/L, fluoride: 0.3 mg/L, 2.3 ppm $\textrm{CaCO}_3$) gives a tip angle $\alpha=123.8\pm 0.9^\circ$, an average difference of nearly $3^\circ$ compared to pure water.
Although the tip angle cannot directly reveal the nature of solutes, we believe that these findings can be used for a quick quality assessment of water purity with an optical device, which could be fully automatized.
As the tip angles have a certain distribution, a quality assessment protocol would require to perform a statistical analysis of the sample.
These results also indicate the significance of the water purity in freezing studies.

Further work must be devoted to the understanding of the effect of solutes and particles on the ice structure and morphology, which is particularly challenging \cite{Dedovets2018}.
This observation would also deserve to be expanded to different freezing geometries such as cold surfaces impinged by drops \cite{Ghabache2016a,Thievenaz2019}.

%%%%%%%%%%%%%%%%%%%%%%%%%%%
\section*{Supplementary Material}
%%%%%%%%%%%%%%%%%%%%%%%%%%%

See supplementary material for the source code used for the image processing.

%%%%%%%%%%%%%%%%%%%%%%%%%%%
\section*{Acknowledgments}
%%%%%%%%%%%%%%%%%%%%%%%%%%%
We acknowledge Charlotte Veillon, Johanna Angibault-Gonzales, and Paul Teixeiera for preliminary experiments.
We thank Dominique Langevin for useful discussions, Mélanie Decraene and Claire Goldmann for assistance.

\bibliography{biblio}

\bibliographystyle{unsrt}
\end{document}